\documentclass[thesis,tocnosub,noragright,centerchapter,12pt]{uiucecethesis09}

\usepackage{tikz}
\usetikzlibrary{positioning}
\usepackage{pgfplots}

\usepackage{titlesec}

\titlespacing*{\subsection}
  {0pt}      
  {*}        
  {0em}    

\usepackage{booktabs}

\usepackage[table]{xcolor}

\makeatletter

\renewcommand{\l@section}{\@dottedtocline{1}{0em}{1.5em}}

\usepackage{setspace}
\usepackage{epsfig}  
\usepackage{graphicx}  
\usepackage{amsmath}  
\usepackage{amssymb}  
\usepackage{lscape}  
\usepackage[justification=raggedright]{caption}	
\usepackage{hyperref}
\usepackage{subcaption} 
\usepackage{algorithm}
\usepackage{array}
\usepackage{algpseudocode}

\msthesis

\title{Enhancing Music Recommendation with User Mood Input}
\author{Terence Zeng}
\department{Electrical and Computer Engineering}
\degreeyear{2025}

\advisor{Abhishek Umrawal}

\renewcommand\thesection{\arabic{section}}

\@addtoreset{section}{chapter} 
\makeatother

\usepackage{xcolor}

\begin{document}

%

%
\maketitle

\parindent 1em%

\frontmatter

%
\begin{abstract}

Recommendation systems have become essential in modern music streaming platforms, due to the vast amount of content available. A common approach in recommendation systems is collaborative filtering, which suggests content to users based on the preferences of others with similar patterns. However, this method performs poorly in domains where interactions are sparse, such as music. Content-based filtering is an alternative approach that examines the qualities of the items themselves. Prior work has explored a range of content-filtering techniques for music, including genre classification, instrument detection, and lyrics analysis. In the literature review component of this work, we examine these methods in detail. Music emotion recognition is a type of content-based filtering that is less explored but has significant potential. Since a user's emotional state influences their musical choices, incorporating user mood into recommendation systems is an alternative way to personalize the listening experience. In this study, we explore a mood-assisted recommendation system that suggests songs based on the desired mood using the energy-valence spectrum. Single-blind experiments are conducted, in which participants are presented with two recommendations --- one generated from a mood-assisted recommendation system and one from a baseline system --- and are asked to rate them. Results show that integrating user mood leads to a statistically significant improvement in recommendation quality, highlighting the potential of such approaches.
\\\\
Subject Keywords: Recommendation systems, music recommendation, mood detection, content-based filtering, information retrieval
\end{abstract}

\begin{acknowledgments}
First, I would like to express my gratitude to my advisor, Professor Abhishek Umrawal, for his invaluable guidance throughout this journey. I am also grateful to my family for their continual encouragement. Lastly, I would like to thank all participants for contributing to this study.
\end{acknowledgments}

%

\setlength{\parskip}{0pt}
\tableofcontents
\setlength{\parskip}{1em}

\mainmatter

%

\section{Introduction}

Recommendations have played a central role in helping people discover new content for centuries. Traditionally, recommendations were made through word-of-mouth - a friend might suggest a good book they recently read, or recommend a new restaurant they dined at. Nowadays, however, the sheer abundance of available content in various media has made it impractical for users to rely solely on personal recommendations. One such medium is music. Spotify, the largest streaming service, has over 100 million songs. Users will struggle to manually discover new songs that are aligned with their preferences. To address this, music streaming services employ recommender systems, which suggest songs that users are likely to enjoy.

Before recommender systems became widely adopted, information retrieval served as a way to filter out vast amounts of content. Spearheaded largely by the foundational work of Rijsbergen \cite{Rijsbergen}, these systems became widely used in areas such as library science, where the challenging of indexing and accessing large amounts of content required more efficient methods.

Recommender systems evolved from information retrieval systems as the demand for personalized access to information grew. While information retrieval systems rely on user queries to search for content, recommender systems offer relevant content by utilizing user behavior, eliminating the need for these queries.

The earliest recommender systems relied primarily on collaborative filtering \cite{Goldberg}, a technique that generates recommendations based on user-interaction data. Collaborative filtering works by identifying users with similar listening patterns and recommending items that those similar users enjoyed. This method saw widespread adoption by the early 2000s on platforms such as Amazon \cite{Linden} due to its simplicity and effectiveness. However, collaborative filtering has several limitations, particularly in the domain of music.

One such limitation is that user-interaction data is sparse. For large song databases, the sparsity can surpass 99.9\% \cite{Dror}, meaning that most users will have listened to less than 0.1\% of all songs. It becomes difficult for recommendation systems to accurately model user preferences when there is so little data about what they have or have not listened to. This sparsity not only limits the system's ability to find personalized recommendations but also reinforces popularity bias, as the system may disproportionately recommend widely-streamed tracks simply because they have more interaction data. This makes it especially difficult for lesser-known artists to gain visibility, ultimately reducing the diversity of recommendations.

This problem is exacerbated by the fact that explicit feedback is rarely provided in music. In other media, such as movies, a numerical rating is assigned by a user to an item. Music streaming services generally lack this functionality - the main types of feedback are a user's listening history and playlists. In fact, even this implicit data isn't reliable - there are many reasons why someone may listen to a song they don't completely enjoy. Therefore, we decided to explore content-based filtering, which examines audio characteristics rather than user-item interactions.

\vspace{1em}

\begin{figure}[h!]
\centering
\begin{tikzpicture}[
    box/.style={rectangle, draw=gray!60, fill=gray!5, thick, minimum width=3.5cm, minimum height=1cm, align=center, rounded corners},
    every node/.style={font=\scriptsize},
    node distance=2.5cm
]

\node[box] (ir) {Information Retrieval};
\node[box, right=0.4cm of ir] (audio) {Audio Analysis};
\node[box, right=0.4cm of audio] (music) {Music Similarity};

\node[above=0.2cm of ir] {Rijsbergen (1979) \cite{Rijsbergen}};
\node[above=0.2cm of audio] {Tzanetakis \& Cook (2000) \cite{Tzanetakis}};
\node[above=0.2cm of music] {Logan \& Salomon (2001) \cite{Logan}};

\draw[thick, yellow!60, fill=yellow!5] (5,-3.5) ellipse (4.5cm and 2cm);
\draw[thick, orange!60, fill=orange!5] (5,-4.25) ellipse (2.5cm and 1.25cm);
\draw[thick, cyan!60, fill=cyan!5] (5,-4.75) ellipse (1.5cm and 0.75cm);

\node[text width=4cm, align=center] at (5, -2.25) {Multimedia \\ (video, audio, text)};
\node at (5, -3.5) {Audio};
\node at (5, -4.75) {Music};
\node at (-3, 0) {\textbf{Methods}};
\node at (-1, -3.5) {\textbf{Domains}};

\draw[->, thick] (ir) -- (audio);
\draw[->, thick] (audio) -- (music);

\draw[->, thick] (ir) -- (1.4, -2.3);
\draw[->, thick] (3, -0.5) -- (3, -3.5);
\draw[->, thick] (music) -- (6.3, -4.4);

\end{tikzpicture}
\caption{Timeline of the development of music similarity analysis.}
\end{figure}
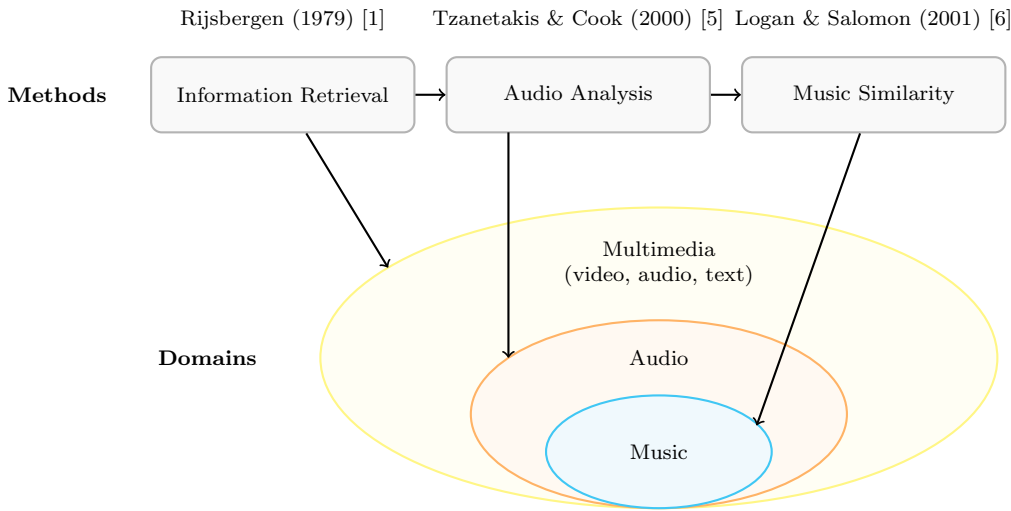

\vspace{1em}

One of the earliest frameworks for audio analysis was MARSYAS, developed by Tzanetakis \& Cook \cite{Tzanetakis}. Inspired by the aforementioned information retrieval techniques, the MARSYAS framework audio extracts some information from the raw signals and stores it in a structured text format, allowing for traditional retrieval techniques to be used. A key innovation of MARSYAS was its approach to audio representation: rather than storing audio as raw waveform data, it transformed it into the frequency domain, giving access to powerful features such as spectral centroid and Mel-frequency Cepstral Coefficients (MFCCs). Some of these features were proven to be effective by the work of Davis \& Mermelstein \cite{davis}. These features were then used in supervised and unsupervised learning models to perform tasks like emotion recognition, genre classification, and instrument detection. By bridging ideas from information retrieval and audio signal processing, MARSYAS laid the foundation for many future content-based music recommendation systems.

Logan \& Salomon \cite{Logan} further explored MFCCs, using them to filter music based on content. Their work was motivated by the growing need to create playlists with similar-sounding songs, since digital media consumption was moving from albums to individual tracks. To find similar-sounding songs, they used frequency analysis combined with k-means clustering within each song. These clusters represent the musical shape of each song. With this system, the proportion of similar-sounding songs was much higher than with a randomly generated playlist. These methods and results showed that audio analysis was an area worth looking at and paved the way for more sophisticated methods.

\subsection{Motivation}

In recent years, as music streaming platforms have scaled to host tens of millions of songs, the need for
effective and personalized recommendation systems has become increasingly central. This has driven a shift
from early collaborative filtering approaches to more sophisticated, content-aware techniques that leverage audio signals, lyrics, emotional context, and user demographics. 

Out of the several content filtering methods discussed in the literature review section, we chose to explore mood-based recommendation based on several factors:
\begin{itemize}
    \item With recent advances in deep learning, signal processing, and natural language processing, there is high potential for progression in this area
    \item It is relatively unexplored compared to other content-filtering techniques
    \item Current trends point to the increase of collection of user data, including relevant contextual data from which we can infer a user's mood
\end{itemize}

\subsection{Contribution}

This thesis contributes to the growing area of affective recommendation systems by developing, implementing, and experimentally evaluating a framework for mood-assisted music recommendations. The primary contribution lies in demonstrating that explicit user mood input, combined with valence–energy audio features, can improve the perceived quality of music recommendations when compared to similarity-based methods alone. 

In addition, this work delivers a functioning web-based platform that integrates multiple external data sources — Spotify, Last.fm, and ReccoBeats — and performs real-time recommendation generation. This system can be used not only for the experiments conducted in this thesis, but also as a foundation for future research about personalized recommendation strategies, human-computer interfaces, or multimodal user input.

\subsection{Organization}

The remainder of this thesis is structured as follows. Section~\ref{sec:methods} describes methods used in our model to quantify results and produce recommendations. Section~\ref{sec:experiments} describes the development of the experimental platform, the APIs and tools used, and the design of the mood-assisted recommendation framework. We also outline the experimental process, including the methods used for comparing baseline and mood-aware recommendations. Section~\ref{sec:results} presents the outcomes of the user study, summarizing trends in participant ratings and comparing performance across both the control and experimental groups. We also interpret these findings, discussing their implications, limitations, and potential contributions to the broader field. Finally, Section~\ref{sec:conclusion} offers concluding remarks and highlights areas for future work.

\subsection{Literature Review}

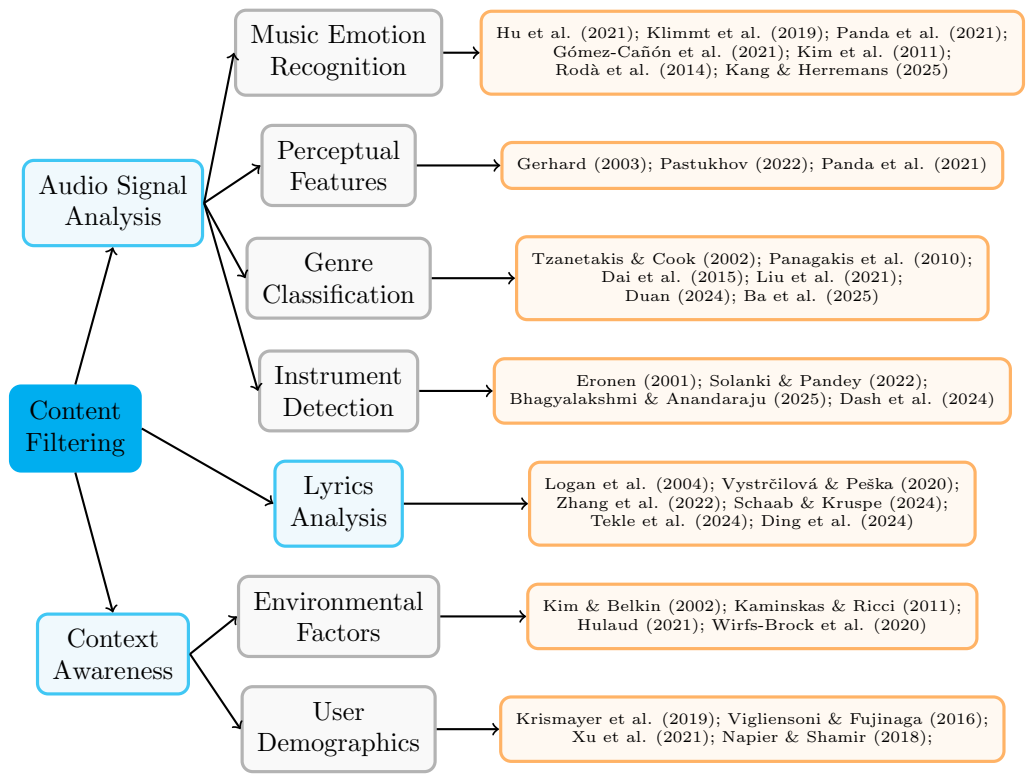
\begin{figure*}[h!]
    \centering
\begin{tikzpicture}[
Title/.style={rectangle, draw=cyan!100, fill=cyan!100, very thick, inner sep=2mm, align=center, rounded corners, font=\footnotesize},
Heading/.style={rectangle, draw=cyan!60, fill=cyan!5, very thick, inner sep=2mm, align=center, rounded corners, font=\footnotesize},
Normal/.style={rectangle, draw=gray!60, fill=gray!5, very thick, inner sep=2mm, align=center, rounded corners, font=\footnotesize},
Citation/.style={rectangle, draw=orange!60, fill=orange!5, very thick, inner sep=2mm, align=center, rounded corners, font=\tiny}
]
\node[Title] (ContentFiltering) at (0.5, 0) {Content \\ Filtering};
\node[Heading] (Audio) at (1, 3) {Audio Signal \\ Analysis};
\node[Heading] (Lyrics) at (4, -1) {Lyrics \\ Analysis};
\node[Heading] (Context) at (1, -3) {Context \\ Awareness};

\node[Normal] (MER) at (4, 5) {Music Emotion \\ Recognition};
\node[Normal] (Perceptual) at (4, 3.5) {Perceptual \\ Features};
\node[Normal] (Genre) at (4, 2) {Genre \\ Classification};
\node[Normal] (Instrument) at (4, 0.5) {Instrument \\ Detection};
\node[Normal] (Environmental) at (4, -2.5) {Environmental \\ Factors};
\node[Normal] (Demographics) at (4, -4) {User \\ Demographics};

\node[Citation] (MER_Cite) at (9.5, 5) {Hu et al. (2021); Klimmt et al. (2019); Panda et al. (2021);\\
Gómez-Cañón et al. (2021); Kim et al. (2011);\\
Rodà et al. (2014); Kang \& Herremans (2025)};
\node[Citation] (Perceptual_Cite) at (9.5, 3.5) {Gerhard (2003); Pastukhov (2022); Panda et al. (2021)};
\node[Citation] (Genre_Cite) at (9.5, 2) {Tzanetakis \& Cook (2002); Panagakis et al. (2010);\\
Dai et al. (2015); Liu et al. (2021);\\
Duan (2024); Ba et al. (2025)};
\node[Citation] (Instrument_Cite) at (9.5, 0.5) {Eronen (2001); Solanki \& Pandey (2022);\\
Bhagyalakshmi \& Anandaraju (2025); Dash et al. (2024)};
\node[Citation] (Lyrics_Cite) at (9.5, -1) {Logan et al. (2004); Vystrčilová \& Peška (2020);\\
Zhang et al. (2022); Schaab \& Kruspe (2024);\\
Tekle et al. (2024); Ding et al. (2024)};
\node[Citation] (Environmental_Cite) at (9.5, -2.5)
{Kim \& Belkin (2002); Kaminskas \& Ricci (2011);\\
Hulaud (2021); Wirfs-Brock et al. (2020)};

\node[Citation] (Demographics_Cite) at (9.5, -4) {Krismayer et al. (2019); Vigliensoni \& Fujinaga (2016);\\
Xu et al. (2021); Napier \& Shamir (2018);};

\draw[->, thick] (ContentFiltering.north) to (Audio.south);
\draw[->, thick] (ContentFiltering.east) to (Lyrics.west);
\draw[->, thick] (ContentFiltering.south) to (Context.north);

\draw[->, thick] (Audio.east) to (MER.west);
\draw[->, thick] (Audio.east) to (Perceptual.west);
\draw[->, thick] (Audio.east) to (Genre.west);
\draw[->, thick] (Audio.east) to (Instrument.west);

\draw[->, thick] (Context.east) to (Environmental.west);
\draw[->, thick] (Context.east) to (Demographics.west);

\draw[->, thick] (Lyrics.east) to (Lyrics_Cite.west);

\draw[->, thick] (MER.east) to (MER_Cite.west);
\draw[->, thick] (Perceptual.east) to (Perceptual_Cite.west);
\draw[->, thick] (Genre.east) to (Genre_Cite.west);
\draw[->, thick] (Instrument.east) to (Instrument_Cite.west);
\draw[->, thick] (Environmental.east) to (Environmental_Cite.west);
\draw[->, thick] (Demographics.east) to (Demographics_Cite.west);

\end{tikzpicture}
\caption{An overview of the most common methods used in content filtering and the respective papers.}
\label{fig:topic-graph}
\end{figure*}

As part of this thesis, we conducted an extensive literature review \cite{zeng2025content} discussing the various types of content filtering and their history. The topics covered are shown in the diagram in Figure \ref{fig:topic-graph}.

\section{Methods}
\label{sec:methods}

In this section, we explore a method for quantifying mood in a way that is pertinent to music. We then propose two models for song selection based on mood characteristics, comparing and contrasting them.

\subsection{Energy-Valence Model}

The Energy-Valence Model, proposed by Russell \cite{russell}, is a way to quantitatively define human mood on two dimensions. Energy (sometimes called arousal) is a measure of how alert or stimulated someone feels. Valence is a measure of how pleasant an emotion is. Both of these values range from 0 to 1.

\vspace{1em}

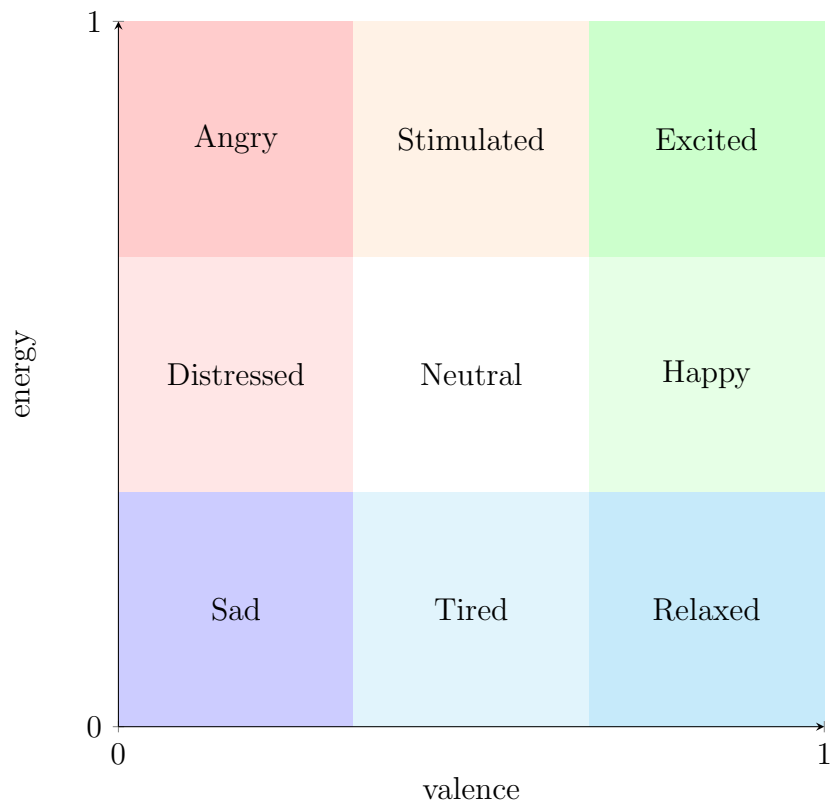
\begin{figure}[!htbp]
    \centering
    \begin{tikzpicture}
    \begin{axis}
    [axis lines=left, xlabel=valence, ylabel=energy,
    xmin = 0, xmax = 1, ymin = 0, ymax = 1,
    xtick = {0, 1}, ytick = {0, 1},
    width=0.8*\textwidth, height=0.8*\textwidth,
    axis on top=true]

    \fill[blue!20]      (axis cs:0,0) rectangle (axis cs:1/3,1/3);
    \fill[cyan!10]      (axis cs:1/3,0) rectangle (axis cs:2/3,1/3);
    \fill[cyan!20]      (axis cs:2/3,0) rectangle (axis cs:1,1/3);
    \fill[red!10]       (axis cs:0,1/3) rectangle (axis cs:1/3,2/3);
    \fill[white!10]     (axis cs:1/3,1/3) rectangle (axis cs:2/3,2/3);
    \fill[green!10]     (axis cs:2/3,1/3) rectangle (axis cs:1,2/3);
    \fill[red!20]       (axis cs:0,2/3) rectangle (axis cs:1/3,1);
    \fill[orange!10]    (axis cs:1/3,2/3) rectangle (axis cs:2/3,1);
    \fill[green!20]     (axis cs:2/3,2/3) rectangle (axis cs:1,1);
    
    \node at (axis cs:1/6, 1/6) {Sad};
    \node at (axis cs:1/6, 1/2) {Distressed};
    \node at (axis cs:1/6, 5/6) {Angry};
    \node at (axis cs:1/2, 5/6) {Stimulated};
    \node at (axis cs:5/6, 5/6) {Excited};
    \node at (axis cs:5/6, 1/2) {Happy};
    \node at (axis cs:5/6, 1/6) {Relaxed};
    \node at (axis cs:1/2, 1/6) {Tired};
    \node at (axis cs:1/2, 1/2) {Neutral};
    \end{axis}
\end{tikzpicture}
    \caption{A mapping of mood categories onto the energy-valence model, where colors indicate different mood regions. Figure adapted from Thayer \cite{thayer1990biopsychology}.}
    \label{fig:placeholder}
\end{figure}

\pagebreak

\begin{figure}[!htbp]
    \centering
    \begin{tikzpicture}
    \begin{axis}
    [axis lines=left, xlabel=valence, ylabel=energy,
    xmin = 0, xmax = 1, ymin = 0, ymax = 1,
    xtick = {0, 1}, ytick = {0, 1},
    width=0.8*\textwidth, height=0.8*\textwidth,
    axis on top=true,
    only marks,
    point meta=explicit symbolic,
    nodes near coords,
    every node near coord/.append style={anchor=west}
    ]

    \fill[blue!20]      (axis cs:0,0) rectangle (axis cs:1/3,1/3);
    \fill[cyan!10]      (axis cs:1/3,0) rectangle (axis cs:2/3,1/3);
    \fill[cyan!20]      (axis cs:2/3,0) rectangle (axis cs:1,1/3);
    \fill[red!10]       (axis cs:0,1/3) rectangle (axis cs:1/3,2/3);
    \fill[white!10]     (axis cs:1/3,1/3) rectangle (axis cs:2/3,2/3);
    \fill[green!10]     (axis cs:2/3,1/3) rectangle (axis cs:1,2/3);
    \fill[red!20]       (axis cs:0,2/3) rectangle (axis cs:1/3,1);
    \fill[orange!10]    (axis cs:1/3,2/3) rectangle (axis cs:2/3,1);
    \fill[green!20]     (axis cs:2/3,2/3) rectangle (axis cs:1,1);

    \addplot [
        mark=*
    ] coordinates {
        (0.10, 0.17) [A]
        (0.08, 0.59) [B]
        (0.21, 0.87) [C]
        (0.82, 0.74) [D]
        (0.42, 0.48) [E]
        (0.51, 0.73) [F]
        (0.71, 0.44) [G]
    };
  
    \end{axis}
\end{tikzpicture}
    \caption{Some examples of popular songs and their placement on the valence-energy plane.}
    \label{fig:song-graph-specific}
\end{figure}
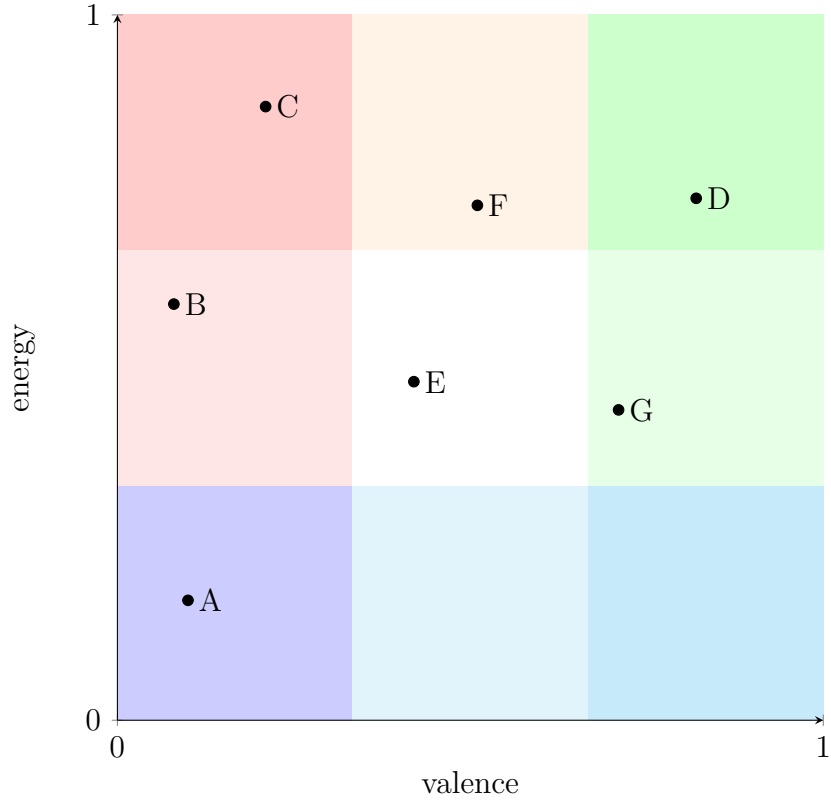

Naturally, songs can also be placed on this plane — valence and energy values can be calculated for any song using signal processing techniques. Figure \ref{fig:song-graph-specific} shows some examples of songs and their position on the plane.

\begin{table}[!htbp]
  \centering
  \caption{Details of songs from Figure \ref{fig:song-graph-specific}}
  \label{tab:example}
  \begin{tabular}{lccc}
    \toprule
    Label & Song & Valence & Energy \\
    \midrule
    A & Last Words of a Shooting Star by Mitski & 0.10 & 0.17 \\
    B & Talking to the Moon by Bruno Mars & 0.08 & 0.59 \\
    C & Numb by Linkin Park & 0.21 & 0.87 \\
    D & Love Me Not by Ravyn Lenae & 0.82 & 0.74 \\
    E & I Wanna Be Yours by Arctic Monkeys & 0.42 & 0.48 \\
    F & NOKIA by Drake & 0.51 & 0.73 \\
    G & I'm Yours by Jason Mraz & 0.71 & 0.44 \\
    \bottomrule
  \end{tabular}
\end{table}

\begin{figure}[!hbtp]
    \centering
    \begin{tikzpicture}
    \begin{axis}
    [axis lines=left, xlabel=valence, ylabel=energy,
    xmin = 0, xmax = 1, ymin = 0, ymax = 1,
    xtick = {0, 1}, ytick = {0, 1},
    width=0.8*\textwidth, height=0.8*\textwidth,
    axis on top=true,
    only marks,
    point meta=explicit symbolic,
    nodes near coords,
    every node near coord/.append style={anchor=west}
    ]

    \fill[blue!20]      (axis cs:0,0) rectangle (axis cs:1/3,1/3);
    \fill[cyan!10]      (axis cs:1/3,0) rectangle (axis cs:2/3,1/3);
    \fill[cyan!20]      (axis cs:2/3,0) rectangle (axis cs:1,1/3);
    \fill[red!10]       (axis cs:0,1/3) rectangle (axis cs:1/3,2/3);
    \fill[white!10]     (axis cs:1/3,1/3) rectangle (axis cs:2/3,2/3);
    \fill[green!10]     (axis cs:2/3,1/3) rectangle (axis cs:1,2/3);
    \fill[red!20]       (axis cs:0,2/3) rectangle (axis cs:1/3,1);
    \fill[orange!10]    (axis cs:1/3,2/3) rectangle (axis cs:2/3,1);
    \fill[green!20]     (axis cs:2/3,2/3) rectangle (axis cs:1,1);

    \addplot [
        mark=*
    ] coordinates {
        (0.33, 0.62)
        (0.24, 0.67)
        (0.24, 0.67)
        (0.37, 0.85)
        (0.22, 0.57)
        (0.28, 0.64)
        (0.81, 0.73)
        (0.45, 0.51)
        (0.41, 0.71)
        (0.59, 0.30)
        (0.42, 0.77)
        (0.38, 0.61)
        (0.31, 0.76)
        (0.10, 0.78)
        (0.71, 0.21)
        (0.35, 0.72)
        (0.41, 0.69)
        (0.53, 0.77)
        (0.10, 0.69)
        (0.42, 0.60)
        (0.26, 0.72)
        (0.27, 0.50)
        (0.42, 0.80)
        (0.20, 0.60)
        (0.15, 0.78)
        (0.46, 0.41)
        (0.43, 0.79)
        (0.30, 0.68)
        (0.33, 0.39)
        (0.80, 0.84)
        (0.35, 0.65)
        (0.57, 0.56)
        (0.55, 0.73)
        (0.22, 0.63)
        (0.24, 0.69)
        (0.50, 0.81)
        (0.27, 0.68)
        (0.04, 0.56)
        (0.32, 0.85)
        (0.47, 0.39)
        (0.26, 0.64)
        (0.46, 0.71)
        (0.48, 0.42)
        (0.15, 0.53)
        (0.59, 0.69)
    };
  
    \end{axis}
\end{tikzpicture}
    \caption{An example distribution of a user's most-listened songs. This particular distribution represents the author's most listened to songs in the past 6 months.}
    \label{fig:song-graph-general}
\end{figure}
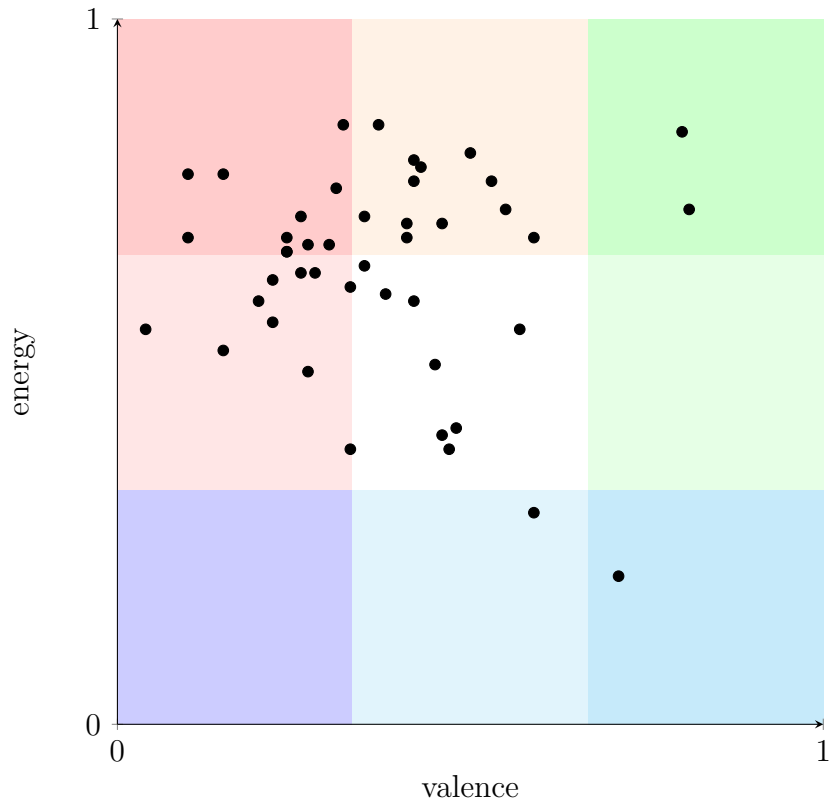

A typical user will listen to songs across most of the entire valence-energy plane, as shown in Figure \ref{fig:song-graph-general}.

\subsection{$k$-Nearest Neighbors}

One method to select songs from a set given a user's mood is to simply select the songs that are closest to a user's mood. Since the user mood and song moods can be mapped as points on a plane, we calculate ``distance'' as the Euclidean distance of a song's valence-energy attributes to those of the user's.

Suppose the user mood is given as some tuple $(v, e)$, where $v = \text{valence} \in [0, 1], e = \text{energy} \in [0, 1]$. Likewise, song attributes are stored in arrays $V$ and $E$.
For every song $i$, we compute that its squared Euclidean distance is $(V[i]-v)^2 + (E[i]-e)^2$. Squared distance is used instead of the actual Euclidean distance because taking a square root does not change the relative ordering of distances, allowing the algorithm to reduce computational cost while preserving the correctness of the ranking. After all distances are computed, the songs are sorted in ascending order of these distances, assuring that the best matches appear first.

\begin{algorithm}[!htbp]
    \caption{$k$-Nearest Neighbors}
    \begin{algorithmic}
        \Require user mood $(v, e)$, \# of recommendations $k$, songs $A[1, 2, \ldots, n]$, valences $V[1, 2, \ldots, n]$, energies $E[1, 2, \ldots, n]$
        \Statex
        \State $D[1, 2, \ldots, n]$    \Comment{array containing distance of each song}
        \ForAll{$i \in [1, n]$}
            \State $D[i] \gets (V[i]-v)^2 + (E[i]-e)^2$
        \EndFor
        \Statex
        \State Sort $A$ based on ascending order of $D$        \Comment{using some lambda sort method}
        \State \Return $A[1, 2, \ldots, k]$
    \end{algorithmic}
\end{algorithm}

\subsection{Softmax/Boltzmann Distribution}

The primary issue with $k$-nearest neighbors is that the input fully determines the output. A good recommendation system should produce varying output, even with the same user mood and song set. This is because the average user's music tastes will change very slowly over time, and they may be in the same moods over and over again. Recommending familiar songs will provide less utility than recommending less optimal but novel ones.

The softmax selection method \cite{softmax} introduces controlled randomness into the recommendation process by converting the distance scores into a probability distribution. It can alternatively be called the Boltzmann method because it utilizes an exponential decay function, as in the Boltzmann probability distribution. Just like before, each song's Euclidean distance from the user's mood, $D[i]$, is computed. Instead of deterministically selecting the closest songs, these distances are transformed into weights $W[i] = \exp\!\left(-D[i]/k\right)$, where the decay factor $k$ determines how sharply the probabilities decrease with distance. The weights are then normalized to obtain a valid probability distribution --- the sum of probabilities should be equal to 1. With this method, we ensure nearby songs are the most likely to be chosen, while introducing some randomness to increase recommendation novelty.

\begin{algorithm}[!htbp]
    \caption{Softmax Song Selection}
    \label{alg:softmax}
    \begin{algorithmic}
        \Require user mood $(v, e)$, decay factor $k$, \# of recommendations $r$, songs $A[1,\ldots,n]$, valences $V[1,\ldots,n]$, energies $E[1,\ldots,n]$
        \Statex
        \State $D[1,\ldots,n]$ \Comment{distance between each song and user mood}
        \ForAll{$i \in [1,n]$}
            \State $D[i] \gets \sqrt{(V[i] - v)^2 + (E[i] - e)^2}$
        \EndFor
        \Statex
        \State $W[1,\ldots,n]$ \Comment{unnormalized Boltzmann weights}
        \ForAll{$i \in [1,n]$}
            \State $W[i] \gets \exp\!\left(-\frac{D[i]}{k}\right)$
        \EndFor
        \Statex
        \State $P[1,\ldots,n]$ \Comment{normalized probabilities}
        \State $Z \gets \sum_{i=1}^{n} W[i]$
        \ForAll{$i \in [1,n]$}
            \State $P[i] \gets W[i] / Z$
        \EndFor
        \Statex
        \State Sample $r$ distinct indices with probabilities $P$ \Comment{using np.random.choice}
        \State \Return corresponding songs from $A$
    \end{algorithmic}
\end{algorithm}

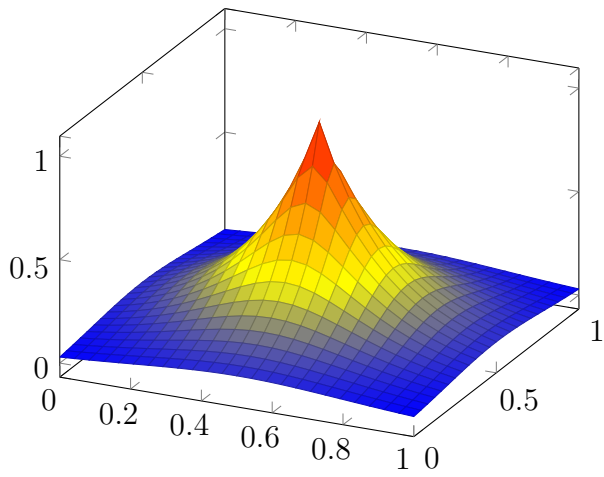
\begin{figure}[!htbp]
    \centering
    \begin{tikzpicture}
      \begin{axis}[domain=0:1,y domain=0:1]
        \addplot3[surf] {exp(-5*sqrt((x-0.5)^2+(y-0.5)^2))};
      \end{axis}
    \end{tikzpicture}
    \caption{An illustration of the (unnormalized) distribution of songs selected, based on their position on the valence-energy plane. Here, the user's target mood is (0.5, 0.5) and the decay factor is $k$ = 5.}
    \label{fig:3d-softmax}
\end{figure}

We can see a 3-dimensional visualization of the Boltzmann distribution in Figure \ref{fig:3d-softmax}, where the $z$-axis represents the (unnormalized) weight of a song. The $x$-axis and $y$-axis represent valence and energy values, respectively.

\section{Experiments}
\label{sec:experiments}

\begin{figure}[!htbp]
    \centering
    \includegraphics[width=\textwidth]{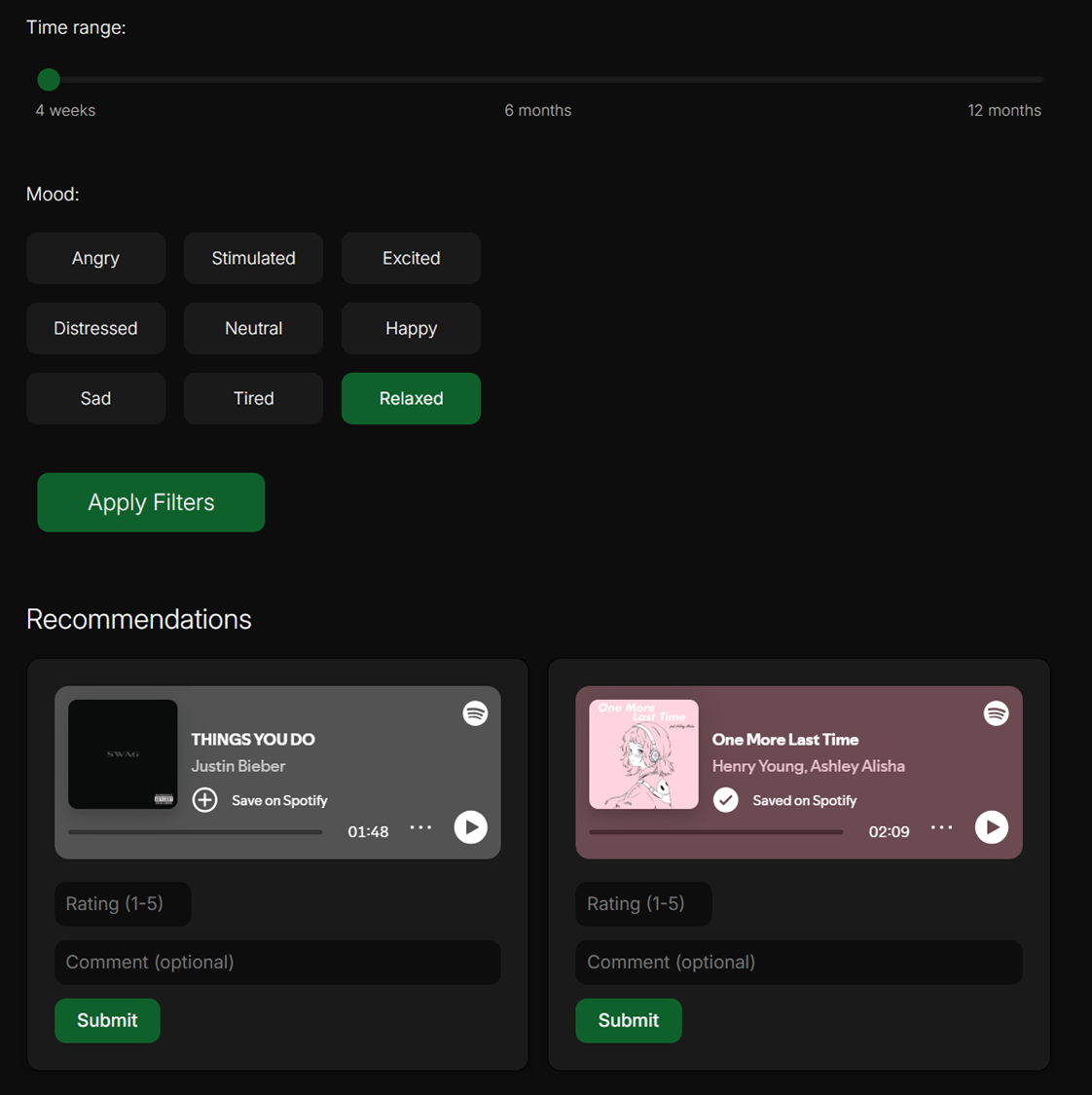}
    \caption{Flowchart outlining the experimental process, including tools and methods used at each stage.}
    \label{fig:moodtune_website}
\end{figure}

\begin{figure}[!htbp]
    \centering
    \includegraphics[width=\textwidth]{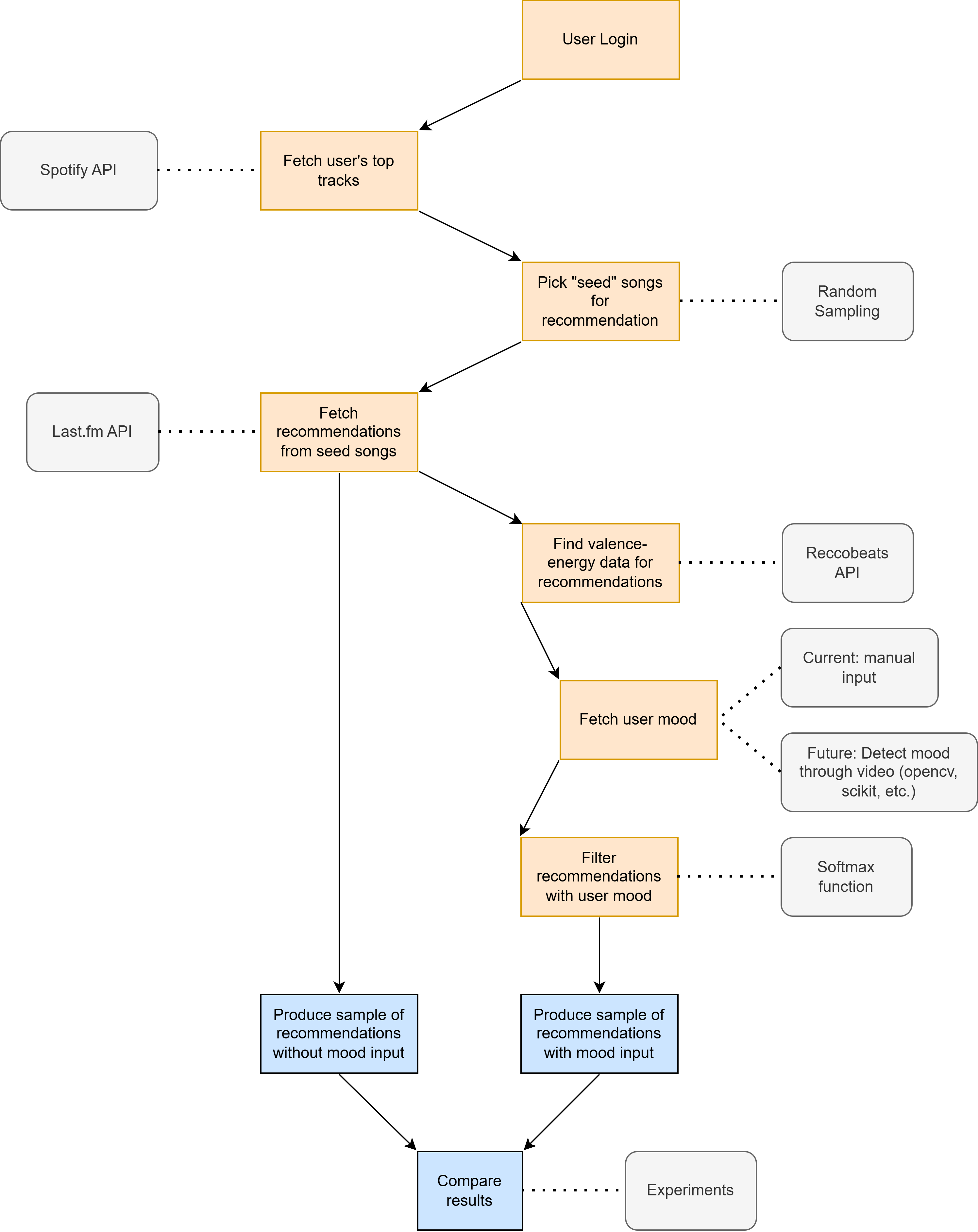}
    \caption{Flowchart outlining the experimental process, including tools and methods used at each stage.}
    \label{fig:methods_flowchart}
\end{figure}

In this section, we describe the experimental flow (Figure \ref{fig:methods_flowchart}) used to evaluate the effectiveness of mood-assisted music recommendation compared to a traditional system. We outline the website platform (Figure \ref{fig:moodtune_website}), data collection procedures, and the structure of the user study, including how mood inputs and recommendations were presented.

\vspace{1em}

\pagebreak

To conduct experiments, we created a website to collect data on user preferences and recommendation ratings. We chose to create a website rather than some simpler form of data collection (such as a Google Form) because we needed a platform capable of interacting with multiple third-party music APIs, processing user mood input, and generating personalized recommendations in real time. To support these requirements, we developed a fully functional web application that integrates Spotify user data, Last.fm recommendation services, and ReccoBeats audio feature analysis. The website served not only as an interface for user participation, but also executes the experimental flow illustrated in Figure~\ref{fig:methods_flowchart}. This section provides a detailed description of the design, implementation, and deployment of the website.

\subsection{Methods}

\subsubsection{Overview of the Website Platform}

On the website's landing page, users first log into their Spotify accounts. Users' top tracks are retrieved through the Spotify Web API. From these tracks, a subset of ``seed songs'' is randomly selected and passed to the Last.fm API to generate similar tracks. For each recommended track, valence and energy information is obtained through the ReccoBeats API. These features enable the filtering and ranking operations that we discussed in the softmax song selection process (Algorithm \ref{alg:softmax}). Finally, users are presented with a pair of recommendations --- one of which is a ``baseline'' or ``control'' recommendation, and one which is filtered with a specific mood target. Users can then give these recommendations ratings and leave feedback in the form of comments. All responses are recorded for later statistical analysis. By collecting ratings from both systems under identical conditions, this platform enables systematic comparison of the effectiveness of mood integration in song recommendations. 

\subsubsection{APIs and Data Integration}

The system integrates data from three APIs, each supporting a different stage of the experimental pipeline.

\begin{itemize}
    \item \textbf{Spotify API:} The Spotify Web API is used to authenticate users and retrieve their top tracks over various time ranges (short-term, medium-term, and long-term). The API provides important metadata for each track, such as its artist and Spotify ID, both of which can be used as identifiers in other music platforms. It can also provide information about a user's public playlists.
    \item \textbf{Last.fm API:} Spotify does not provide an endpoint for retrieving ``similar songs,” which is essential to generating new, unheard recommendations from a user's existing preferences. To address this limitation, the website relies on the Last.fm API. For each seed song, the Last.fm method \texttt{track.getSimilar} is used to produce one or more semantically related tracks. Since Last.fm and Spotify IDs are not one-to-one, we use the Spotify search endpoint to reconcile the Last.fm recommendations with specific Spotify song IDs.
    \item \textbf{Reccobeats API:} The ReccoBeats API provides valence and energy information for individual tracks. These values are used to model emotional characteristics of songs and filter recommendations according to user mood input. For each track, two API queries are performed: one to map Spotify's track ID to a ReccoBeats track identifier, and another to extract the corresponding audio features. These features play a central role in the computation of mood-based probabilities during the recommendation step.
\end{itemize}

\subsection{Website}

\subsubsection{Framework: Flask}

The website backend is implemented entirely in Python using the Flask microframework. Flask was chosen primarily because of its minimal and flexible architecture, suitable for a research prototype. It provides direct control over HTTP routing, session handling, and API integration. Additionally, Flask integrates naturally with Python. Python is the ideal language for implementing the various algorithms used, especially because of the prevalence of packages such as numpy. Lastly, Flask simplifies the incorporation of Jinja templates for rendering results, allowing server-side generation of dynamic pages without requiring a complex frontend framework.

\subsubsection{Multithreading and Asynchronous API Calls}

A significant challenge encountered during development was how to deal with the hundreds of API calls required throughout the process of generating the two recommendations. For each step in the process, the system may request dozens of track lookups, similarity searches, and audio-feature queries across Spotify, Last.fm, and ReccoBeats. Executed sequentially, these requests may take a few minutes, rendering the website nearly impossible to use.

To overcome this limitation, the website makes extensive use of asynchronous programming via Python's \texttt{asyncio} library and the \texttt{aiohttp} client. Rather than issuing API calls one at a time, the system launches many requests concurrently and processes the results as they return. This approach reduces total latency from hundreds of seconds to a few seconds, enabling the website to remain responsive and making the data collection process more efficient for participants.

\subsubsection{Deployment and Hosting on PythonAnywhere}

Currently, due to logistical issues, the website is locally hosted on a single computer. In the near future, we plan to host this website on PythonAnywhere, a cloud hosting platform for Python applications. PythonAnywhere was selected due to its compatibility with Flask applications and its simplicity of deployment.

Deployment requires configuration of a WSGI application entry point, environment variable management, and adjustments to ensure that asynchronous code executes reliably under PythonAnywhere's runtime environment. Additionally, the free tier of PythonAnywhere restricts outbound internet access to non-whitelisted domains, complicating communication with the ReccoBeats API. This limitation required additional testing and workarounds during development.

Despite these challenges, hosting the site on PythonAnywhere will ensure that participants can access the system from any device with a browser, enabling remote data collection for the experiments. The controlled hosting environment also guarantees consistent execution conditions across all user sessions, which is essential for ensuring the reliability and comparability of the collected data.
\section{Results}
\label{sec:results}

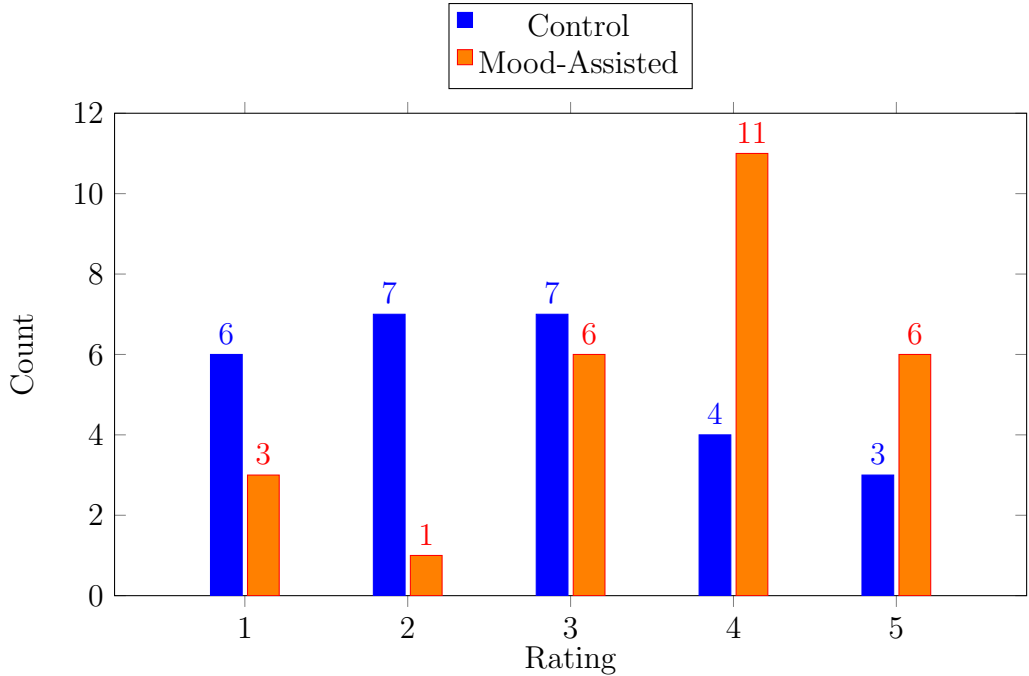
\begin{figure}[!htbp]
\centering
\begin{tikzpicture}
\begin{axis}[
    ybar,
    bar width=12pt,
    width=\textwidth,
    height=8cm,
    ymin=0,
    ymax=12,
    enlarge x limits=0.2,
    legend style={at={(0.5,1.05)},anchor=south,legend columns=1},
    symbolic x coords={1,2,3,4,5},
    xtick=data,
    xlabel={Rating},
    ylabel={Count},
    nodes near coords,
    nodes near coords align={vertical},
    legend image code/.code={
        \draw[#1,fill=#1] (0cm,0cm) rectangle (0.2cm,0.2cm);
    }
]

\addplot+[fill=blue] coordinates {
    (1,6)
    (2,7)
    (3,7)
    (4,4)
    (5,3)
};

\addplot+[fill=orange] coordinates {
    (1,3)
    (2,1)
    (3,6)
    (4,11)
    (5,6)
};

\legend{Control, Mood-Assisted}

\end{axis}
\end{tikzpicture}
\caption{Distribution of participant ratings for the control and mood-assisted recommendation systems across a 5-point scale.}
\label{fig:rating-graph}
\end{figure}

In this section, we present the results of this experiment. Figure \ref{fig:rating-graph} shows all the ratings, comparing those from the control recommendation model and those from the mood-assisted model. Table \ref{table:mood-results} divides the results by the mood chosen.

\vspace{1em}

\subsection{Statistical Analysis}

Across all participants, the mood-assisted recommendations exhibited higher mean ratings than the baseline recommendations. Mood-assisted song recommendations received a mean rating of $3.59$, while control recommendations received a mean rating of $2.67$. The distribution of mood-assisted scores shows an upward shift, with fewer low ratings, compared to the control recommendations.

To quantify the significance of these results, we can use the Mann-Whitney $U$ test. Note that we don't use the $t$-test here because the data may not be normally distributed. 

With the Mann-Whitney $U$ test, we first calculate the rank of each rating (since many songs will have the same ratings):
\begin{itemize}
    \item \textbf{5}: (1+9)/2 = 5
    \item \textbf{4}: (10+24)/2 = 17
    \item \textbf{3}: (25+37)/2 = 31
    \item \textbf{2}: (38+45)/2 = 41.5
    \item \textbf{1}: (46+54)/2 = 50
\end{itemize}

Then, we calculate the two rank sums:
\begin{itemize}
    \item \textbf{Control}: $(3\cdot5 + 4\cdot17 + 7\cdot31 + 7\cdot41.5 + 6\cdot 50) = 890.5$
    \item \textbf{Mood-Assisted}: $(6\cdot5 + 11\cdot17 + 6\cdot31 + 1\cdot41.5 + 3\cdot50) = 594.5$
\end{itemize}

We calculate the standard error:
\begin{align*}
\sigma_{U} &= \sqrt{\frac{n_1\cdot n_2 \cdot (n_1+n_2+1)}{12}} \\
&= \sqrt{\frac{27 \cdot 27 \cdot (27+27+1)}{12}} \\
&= \sqrt{\frac{40095}{12}} \\
&\approx 57.8
\end{align*}

Then, we calculate the $z$-score:
\begin{align*}
    z &= \frac{\text{min rank sum} - \mu_U}{\sigma_U} \\
    &= \frac{594.5 - 742.5}{57.8} \\
    &\approx -2.56
\end{align*}

Converting this $z$-score into a $p$-value results in $\mathbf{p \approx 0.010}$ (assuming two-tailed hypothesis), demonstrating statistical significance.

\subsection{Discussion}

\begin{table}[!htbp]
\centering
\label{tab:mood_ratings}
\begin{tabular}{lcc}
\hline
\textbf{Mood} & \textbf{Control} & \textbf{Mood-Assisted} \\
\hline
\rowcolor{green!40} Relaxed    & 2.33 & 5.00 \\
\rowcolor{green!30} Sad        & 2.00 & 4.00 \\
\rowcolor{green!20} Tired      & 2.33 & 3.67 \\
\rowcolor{green!0} Distressed & 2.67 & 2.67 \\
\rowcolor{green!5} Neutral    & 3.33 & 3.67 \\
\rowcolor{green!5} Happy      & 3.67 & 4.00 \\
\rowcolor{green!15} Angry      & 1.00 & 2.00 \\
\rowcolor{green!0} Stimulated & 4.67 & 4.67 \\
\rowcolor{green!10} Excited    & 2.00 & 2.67 \\
\hline
\end{tabular}
\caption{Mean participant ratings for control and mood-assisted recommendations, grouped by self-reported mood. Row shading intensity corresponds to the magnitude of improvement introduced by mood-assisted recommendations.}
\label{table:mood-results}
\end{table}

The results of the study suggest that incorporating mood as an additional input signal can meaningfully improve the perceived quality of music recommendations. Participants consistently favored tracks selected through the mood-assisted system, indicating that emotion-aligned filtering provides an advantage beyond what is achievable through similarity alone. This finding aligns with previous work suggesting that musical preferences are context-dependent and influenced by emotional state; it provides empirical support for models that incorporate affect as a first-class feature.

Several factors likely contributed to the observed improvements. By mapping user-reported moods onto the valence–energy plane, the system constrained the recommendation space to songs that better matched participants’ emotional expectations. Furthermore, the use of external similarity data (via Last.fm) combined with mood filtering (via ReccoBeats) created a recommendation system that balanced preference continuity and mood relevance. The reduction in low-rated recommendations also suggests that mood filtering may help avoid mismatches that occur when similarity-based models retrieve tracks with undesirable emotional tone.

We can further evaluate the results by analyzing the ratings of each user-selected mood, as in Table \ref{table:mood-results}. Here, it is clear that some mood inputs perform much better than others. For instance, a user wanting to be relaxed will have a significantly improved listening experience, while one who wants to be stimulated may experience no benefit at all.

Of course, this data has its limitations. The reliance on manual mood input introduces a considerable amount of bias. Additionally, this study consists of six participants. The small sample size and single-trial comparison format reduce the generality of the findings. These considerations highlight opportunities for future work, including automated mood detection and larger-scale evaluation.

\section{Conclusion and Future Work}
\label{sec:conclusion}

\subsection{Conclusion}

This thesis investigated whether incorporating user mood can meaningfully enhance the quality of personalized music recommendations. By integrating data from Spotify, Last.fm, and ReccoBeats platforms into a unified experimental platform, the system developed here produced both baseline similarity-based recommendations and biased recommendations aligned with user-reported mood. The results of the user study show that mood-assisted recommendations received higher average ratings and displayed a more favorable distribution of scores, suggesting that emotional context plays a valuable role in shaping musical preference. These findings reinforce the notion that music consumption is not solely determined by long-term taste, but also by short-term emotional states. Incorporating contextual information—whether collected explicitly or inferred from contextual cues—offers a promising path toward more adaptive and personally relevant recommendation systems. In summary, this work provides empirical evidence supporting the integration of mood information in recommendation pipelines, introduces a functional platform for experimentation, and lays the foundation for future developments in emotionally intelligent music recommendation.

\subsection{Technical Limitations}

\begin{itemize}
    \item \textbf{API rate limits:} To speed up the hundreds of API calls that take place in a user interaction, the website employs multithreading to concurrently send API calls. However, making too many API calls in quick succession can overwhelm servers, often leading to rate limiting. Special care was required to maintain the ordering and alignment of results, since asynchronous responses do not necessarily arrive in the same order as their corresponding requests. Even after utilizing various techniques to avoid being rate-limited, users who change settings rapidly in a short time frame may experience crashes on the website. Ideally, these crashes would not occur.
    
    \item \textbf{ID mapping issues:} To manage songs, we use different IDs from Spotify, Last.fm, and Reccobeats. Unfortunately, many songs that are less popular may not be on all three of these platforms, leading to incomplete data. These songs would have been discarded because they are unusable in the recommendation process.
    
    \item \textbf{Undefined audio features:} Though we use the arousal and valence features in this study, they are not clearly defined anywhere. Before it was acquired by Spotify, The Echo Nest was an independent music data platform. This company constructed these features, likely with signal processing and machine learning methods.
    
\end{itemize}

\subsection{Industry Readiness}

The results of this study demonstrate that mood-assisted music recommendation systems are not only conceptually viable but also practically aligned with the needs and constraints of modern music streaming platforms. The proposed framework relies on technologies that are already widely deployed in industry, including user data collection, API calls, web services, and well-established machine learning techniques.

Currently, the pipeline handles requests at a small scale, utilizing asynchronous and parallel API calls to minimize latency. To ensure that the pipeline doesn't fail under more significant loads, paid APIs may be necessary to avoid rate limiting. Additionally, privacy concerns may arise as users log into their accounts and provide contextual data to the music streaming service.

Overall, although the recommendation pipeline shows promise, some technical challenges have to be addressed before it can be used in an industry environment. However, these large companies will likely have the budget to create sophisticated solutions to these technical challenges, and the findings of these experiments suggest that integrating mood-aware personalization into music recommendation can be a worthwhile area to explore.

\subsection{Future Work}

\begin{itemize}
    \item As two dimensions may not capture the full extent of emotions induced by music, a logical next step would be to explore a different and larger number of dimensions, as in \href{https://psycnet.apa.org/fulltext/2008-09984-007.pdf?auth_token=65a5419cd7695e7b64091c24200ed370593bbb42}{Zentner (2008)}
    \item Recent trends point to an increase in multimodal data, such as video and image context. With these developments, user mood can be inferred from visuals, as opposed to simply selected from a few options. The ability to infer mood would create a much stronger experiment, reducing another element of bias.
\end{itemize}

%
\appendix

\backmatter

%
\bibliographystyle{IEEE_ECE}
\bibliography{thesisrefs}


\end{document}